\documentclass[a4paper]{jpconf}
\usepackage{graphicx}
\begin{document}
\title{Appearance of a quark matter phase in hybrid stars}

\author{Tomoki Endo}

\address{Division of Physics, Department of General Education, Kagawa National Colledge of Technology, 355 Chokushi-cho, Takamatsu, Kagawa 761-8058, Japan}

\ead{endo@t.kagawa-nct.ac.jp}

\begin{abstract}
The appearance of quark matter in the core of hybrid stars is a fundamental issue in such compact stars. The central density of these stars is sufficiently high such that nuclear matter undergoes a further change into other exotic phases that consist of hyperons and quarks. However, the equation of state (EOS) for the high-density matter is still not clear and several recent observations have indicated the limitations of the EOSs; theoretical studies should try to elucidate the EOSs. It is believed that the inner regions of the stars should consist of a mixed hadron-quark phase. We study the mixed hadron-quark phase, taking into account finite-size effects, and find that that the mixed phase should be restricted to a narrower region. Therefore, a quark matter phase should appear in the central region.   
\end{abstract}

\section{Introduction}
Currently accepted theories and many experimental results suggest that hadronic matter changes to quark matter in high-density and/or high-temperature regimes by way of the deconfinement transition. The properties of quark matter have been 
 actively studied theoretically in terms of the quark--gluon
 plasma, color superconductivity \cite{alf1,alf3}, magnetism \cite{tat1,tat3},
 and experimentally in terms of relativistic
 heavy-ion collisions \cite{rhic}, and early-universe studies and compact stars
 \cite{mad3,chen}. 
 Such studies are continuing to provide exciting results \cite{risch}.
 Presently, we consider that compact stars consist of not only nuclear matter but also other matter such as hyperons and quarks. We call such stars {\it hybrid stars}.

Because many theoretical calculations have suggested that the deconfinement 
transition is of the first order
at low temperature and high density \cite{pisa,latt},  
 we assume that it is a first-order phase transition here.
 The Gibbs condition \cite{gle1}
 then gives rise to various structured mixed phases (SMPs).
 The SMPs proposed by Heiselberg et al.\ \cite{pet} and Glenndening
 and Pei \cite{gle2} suggest a crystalline structure for the mixed phase
 in the cores of hybrid stars. Such structures are
 called ``droplets'', ``rods'', ``slabs'', ``tubes'', and ``bubbles''. 
 We present the equation of state (EOS) for the mixed phase taking
 into account the charge screening effect \cite{end2} without relying on 
 any approximations. We investigate the inner structures of these stars \cite{end3}. In this paper, we review the inner structures of 
 hybrid stars and apply our EOS to a stationary rotating star.

\section{Formalism and Numerical Results}

We use the EOS given in our paper \cite{end2}, which is presented in our framework.  
Therefore, our approach is only briefly explained here.
Thermal equilibrium is implicitly achieved at $T=0$. We consider that
the hadronic and quark matter and the mixed phase are $\beta$ stable.
We employ density functional theory (DFT) under the local
density approximation \cite{parr,drez}. 
To account for the confinement,
we introduce a sharp boundary between the two phases employing the
bag model \cite{pet,end2} with a surface tension parameter $\sigma$.
The determination of the surface tension between hadronic and quark matter is
a difficult problem.
Thus, many authors have treated
the surface tension as a free parameter and have observed its effect \cite{pet,gle2,alf2}; we take the same approach in this study.
To determine the charge screening effect, we also conduct the calculations 
without the screening effect. 
We then apply the EOS derived in our paper \cite{end2} to
the Tolman--Oppenheimer--Volkoff (TOV) equation \cite{end3}. 
After that, we apply our EOS to a stationary rotating star.
\begin{figure}[htb]
\begin{center}
\includegraphics[width=75mm]{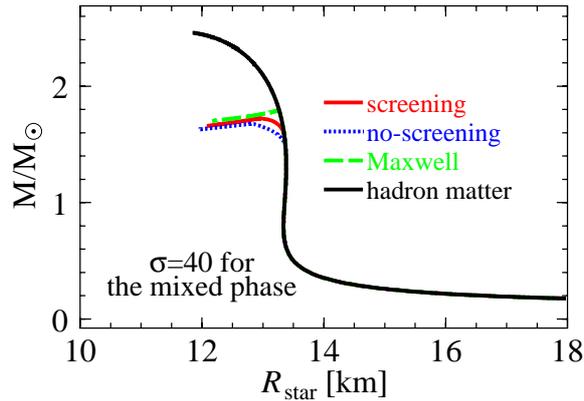}
\caption{ (Color online) Mass--radius relation of stars. The difference between screened and not screened mixed phases is clearly small.}
\label{m-r40}
\end{center}
\end{figure}
\begin{figure}[htb]
\begin{tabular}{cc}
 \begin{minipage}{0.45\hsize}
  \begin{center}
\includegraphics[width=7cm]{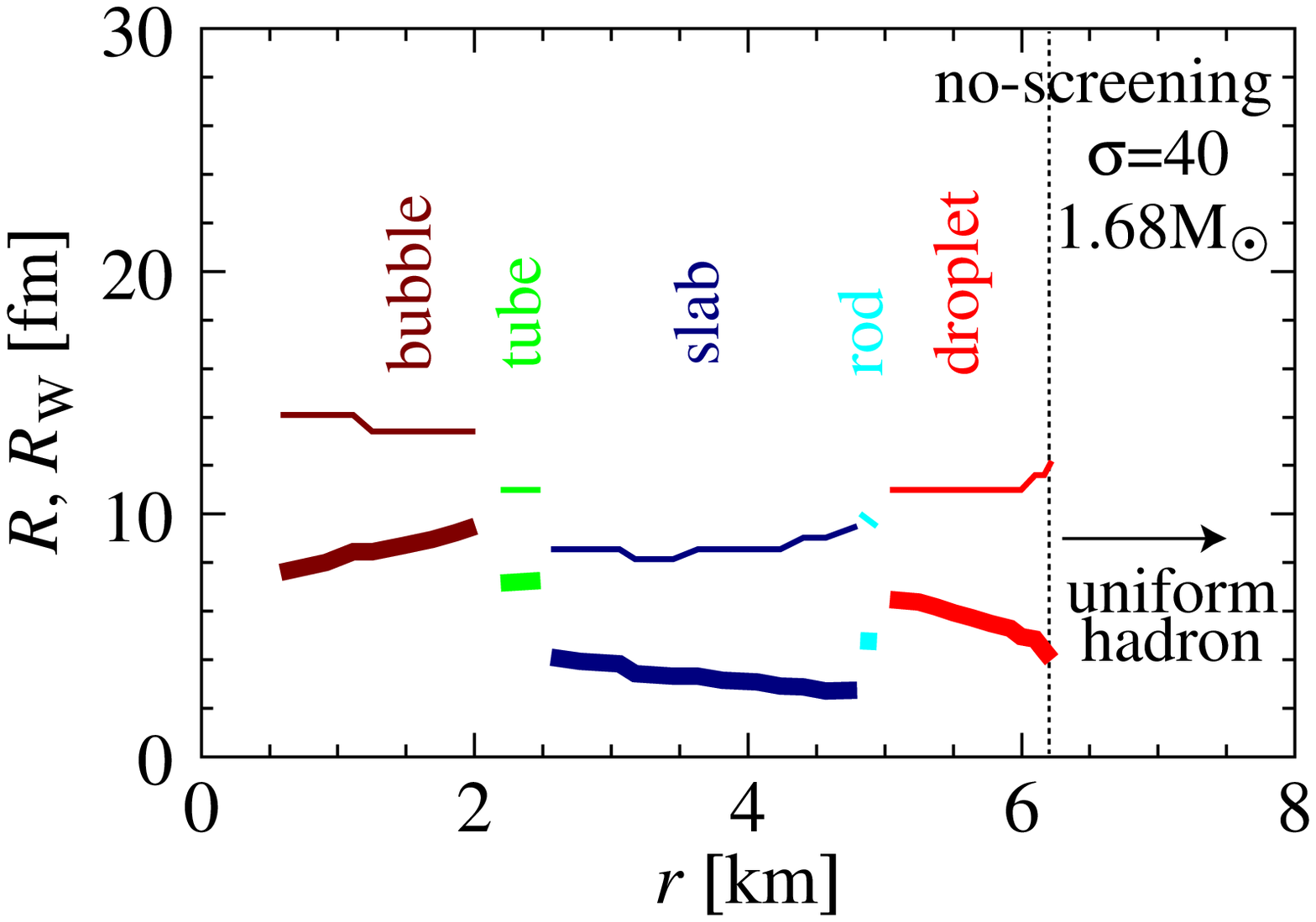}
  \end{center}
\caption{ (Color online) Structure size and cell size in the core region of hybrid
 stars without screening. The radius is 12.6 km. Thick lines denote $R$ and thin lines denote $R_\mathrm{W}$.}
\label{r-ds40no}
\end{minipage}&
\begin{minipage}{0.45\hsize}
  \begin{center}
\includegraphics[width=7cm]{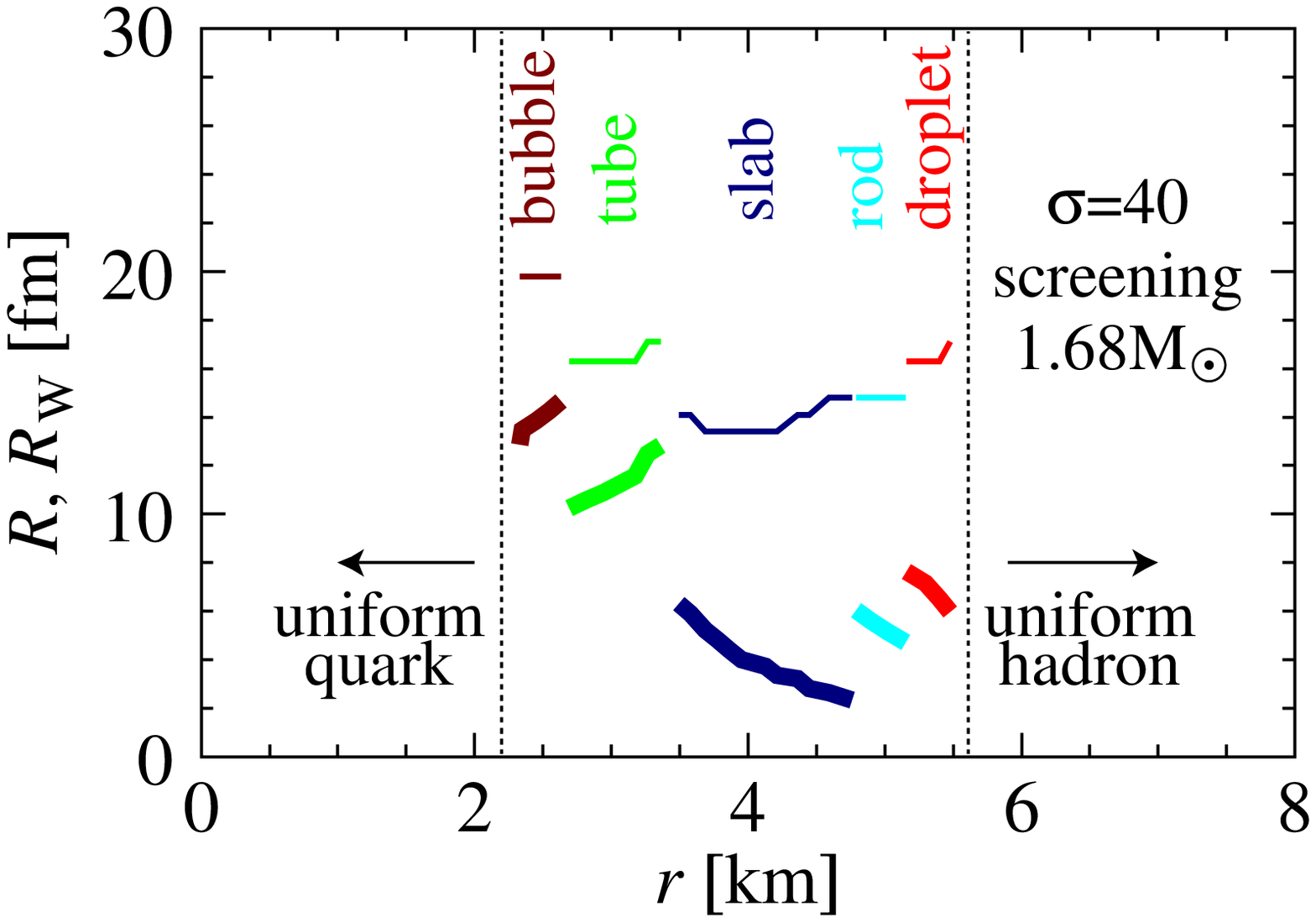}
\end{center}
\caption{ (Color online) Structure size and cell size in the core region of hybrid
 stars with screening. The mixed phase is restricted by charge screening. The radius is 12.9 km.}
\label{r-ds40sc}
 \end{minipage}
\end{tabular}
\end{figure}

Figure\ \ref{m-r40} shows the mass--radius relations of stars with and without screening, using the Maxwell constructrion, and in the case of pure hadronic matter.
There is a slight difference in the region around the maximum mass. However, as this difference is only about $0.05M_\odot$, 
it is small compared with the total mass
of the star. Thus, charge screening 
does not strongly affect the bulk properties of the star.
 On the contrary, we find that the inner structure is greatly
affected by charge screening. Figure\ \ref{r-ds40no} shows the inner core region of a hybrid star without
screening. The mixed phase appears in a large part of the star
and we hardly see the quark matter phase. 
In the case of screening shown in Fig.\ \ref{r-ds40sc}, on the other hand,  
the mixed phase region clearly
becomes narrow and there is a quark matter phase in the central region because of
charge screening.
 Our results suggest that quark matter could exist in the inner regions of compact stars. It is thus possible
to attribute the magnetism of compact stars to
spin polarization \cite{tat4}.
Many theoretical studies have used other models for
the glitch phenomenon \cite{bejg} and gravitational waves \cite{mini}. While we cannot simply apply our EOS to
studying these phenomena, it is 
interesting to compare our results with those of other studies.

\begin{figure}[htb]
\begin{tabular}{cc}
 \begin{minipage}{0.45\hsize}
  \begin{center}
\includegraphics[width=7cm]{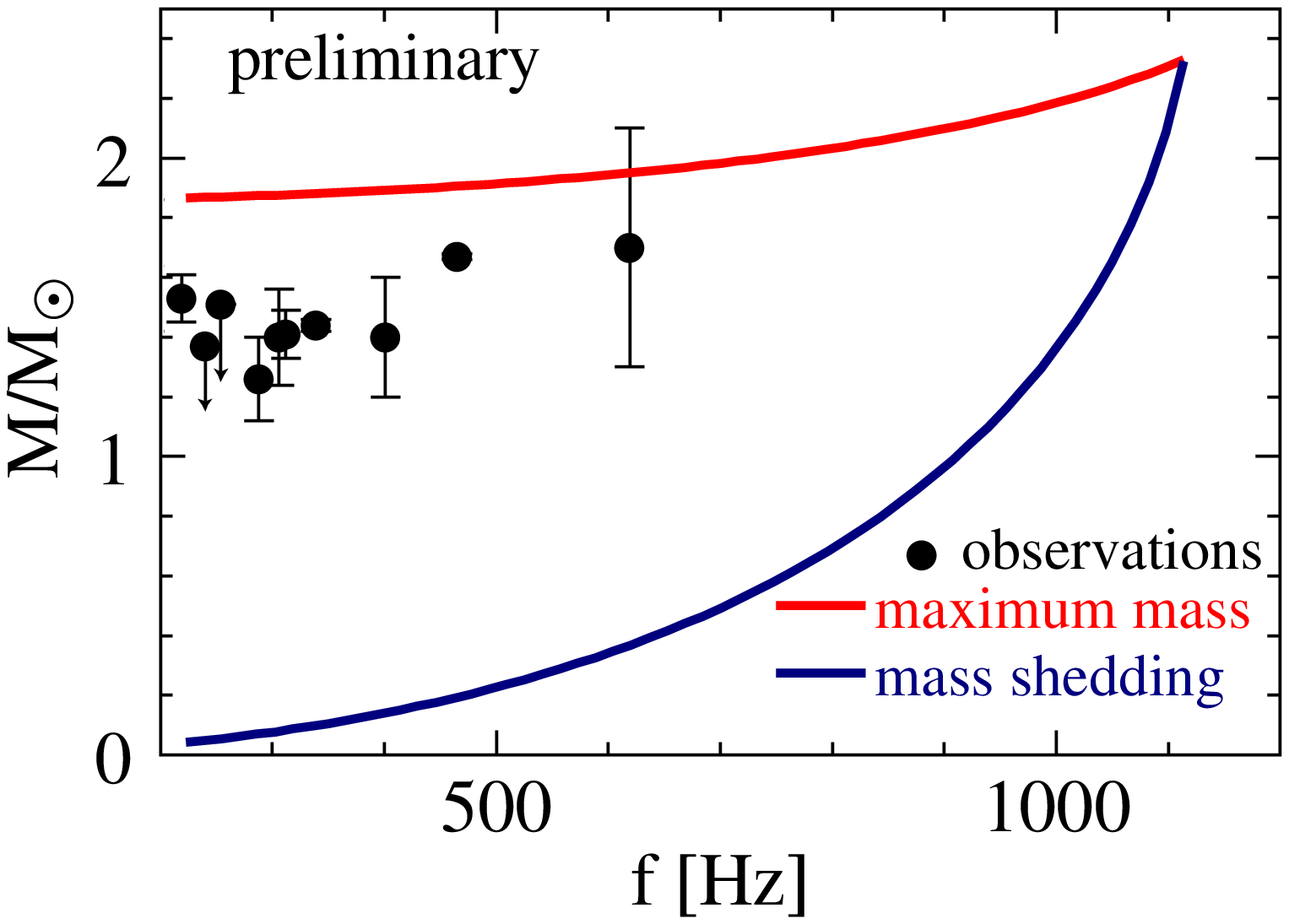}
  \end{center}
\caption{  (Color online) Mass--frequency relation with our model (charge screening) plotted against the observational data listed in \cite{kurk}.}
\label{m-f_sc}
\end{minipage}&
\begin{minipage}{0.45\hsize}
  \begin{center}
\includegraphics[width=7cm]{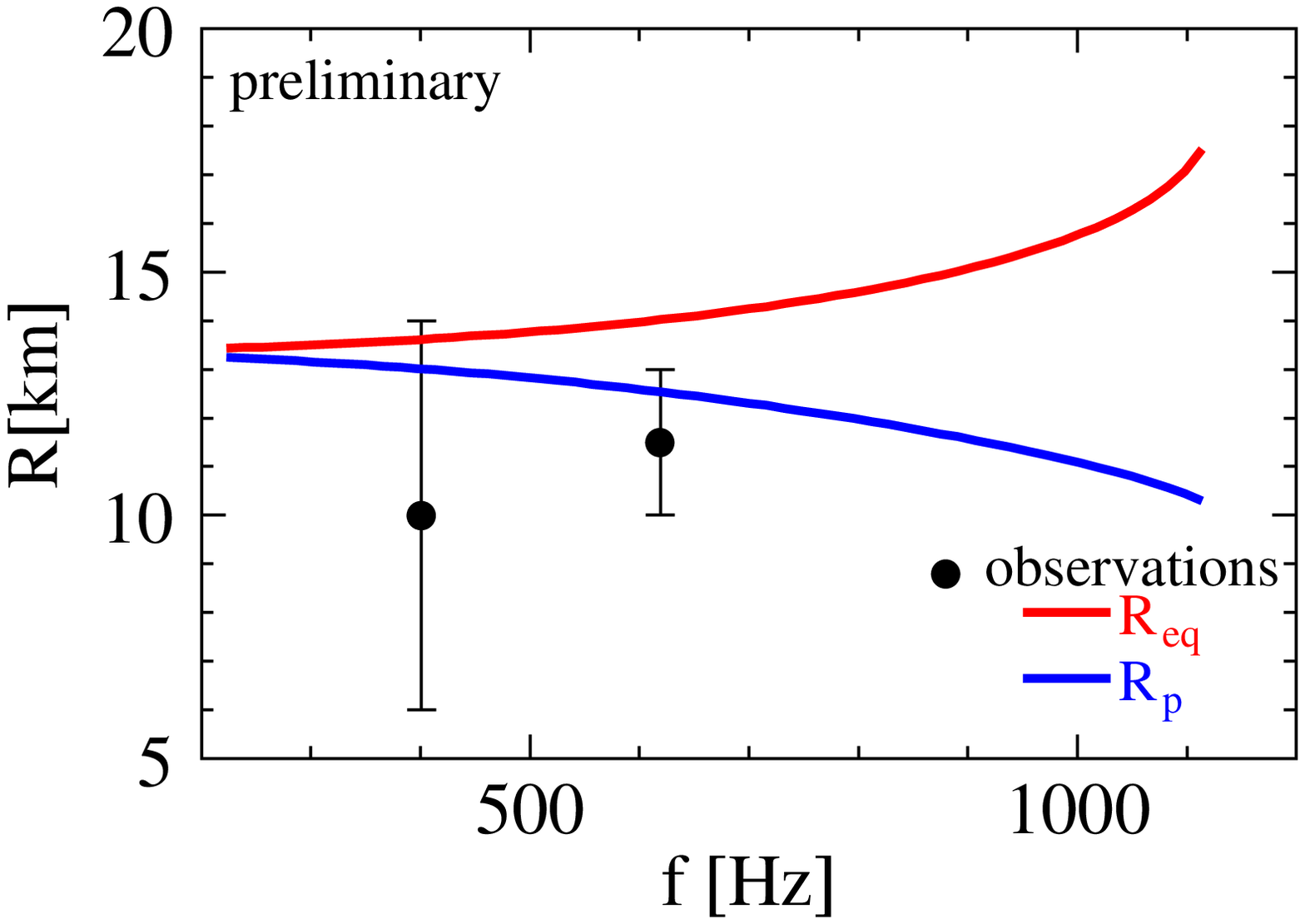}
\end{center}
\caption{ (Color online) Radius--frequency relation of our model plotted against observational data (SAXJ1808.4-3658 and 4U1608-52).}
\label{r-f_sc}
 \end{minipage}
\end{tabular}
\end{figure}

Recently, many theoretical studies tried to take into account the effect of rotation \cite{kurk, orsa, webe,belv}.
We accordingly also try to take into account the effects of rotation in our study. However, rotation in general relativity is very difficult. Therefore, we assume: 1. Stationary rigid rotation (``uniform rotation''); 2. Axial symmetry with respect to the spin axis; and 3. The matter is a perfect fluid.
There is a review of stationary rotation in general relativity \cite{ster} and in other papers \cite{kurk}; we follow their calculation. Then, we apply our EOS to the stationary rotating star.
Figure\ \ref{m-f_sc} shows the result for a rotating star with screening using our EOS. The red curve shows the maximum mass of the star and the blue curve shows the mass-shedding curve, which corresponds to the ``Kepler frequency". The Kepler frequency occurs when the centrifugal force is equal to gravity. Therefore, the right-hand side of the blue curve is not valid. If the red curve is lower than the observations, the EOS should be ruled out. Fortunately, our EOS is consistent with these observations. 
We see an important relation between radius and rotation. Ordinarily, the ``radius'' of the star is single-valued because we consider that the star is spherical. However, if the star is rapidly rotating, we have to pay attention to the different ``radii''.
Because of the effect of the rotation, a star deforms from a sphere to an ellipse. Therefore, we introduce two values, $R_{\mathrm eq}$ and $R_{\mathrm p}$, which are the ``equatorial radius" and the ``polar radius", respectively. Figure\ \ref{r-f_sc} shows $R_{\mathrm eq}$ and $R_{\mathrm p}$ with respect to rotation. If the rotation rate is 400 Hz or faster, the two radii are different. Therefore, we have to note the effects of rotation on rapidly rotating stars.

\section{Summary and Concluding Remarks} \indent

In this study, we demonstrated how charge screening 
affects the hadron--quark mixed phase in the cores of hybrid stars, taking into account rotation effects.
We found that the inner structures are strongly affected.
In particular, a core consisting of quark matter could appear due to
the charge screening effect. Another case, kaon condensation, has been studied \cite{maru} and the results are similar to those of our papers \cite{end2,maru2}. 
We used a simple model for quark matter and nuclear matter.
To obtain a more realistic picture of the hadron--quark phase transition, we need to
take into account color superconductivity \cite{alf1,alf2} and the relativistic mean field theory \cite{shen}. We will then be able to provide more realistic results. 
 Neutron stars have other important physics -- magnetic fields. However, the origin of these magnetic fields is still unknown. There are ways to explain magnetic fields based on the spin-polarization of the quark matter \cite{tat3,tat4}. However, whether the quark matter exists or not strongly depends on the EOS.
In this calculation, we did not take into account magnetic fields. If we include a magnetic field, the resluts are very interesting with respect to the rotation of the star.

\section*{Acknowledgments} \indent

This work was supported in part by the Principal Grant of the Kagawa National College of Technology.

\end{document}